\documentclass[10pt]{article}
\usepackage{fullpage}
\usepackage{amsmath}
\usepackage{amssymb}
\usepackage{amsfonts}
\usepackage{comment}
\usepackage{color}
\usepackage{cite}
\usepackage{subcaption}
\usepackage{graphicx}

\def\##1{{\bf{#1}}}
\def\=#1{\underline{\underline{#1}}}

\def\+
#1{\underline{\bf #1}}
\def\*#1{\underline{\underline{\bf #1}}}

\def\r#1{(\ref{#1})}
\def\l#1{\label{#1}}
\def\c#1{\cite{#1}}

\def\le{\left(}
\def\ri{\right)}
\def\les{\left[}
\def\ris{\right]}
\def\lec{\left\{}
\def\ric{\right\}}

\def\.{\mbox{ \tiny{$^\bullet$} }}

\def\eps{\varepsilon}

\def\epso{\eps_{\scriptscriptstyle 0}}

\def\muo{\mu_{\scriptscriptstyle 0}}

\def\ko{k_{\scriptscriptstyle 0}}

\def\ux{\hat{\#u}_1}
\def\uy{\hat{\#u}_2}
\def\uz{\hat{\#u}_3}

\def\ua{\hat{\#a}}
\def\ub{\hat{\#b}}
\def\uc{\hat{\#c}}

\def\Er{\#E(\#r,\omega)}
\def\Hr{\#H(\#r,\omega)}
\def\DDr{\#D(\#r,\omega)}
\def\Br{\#B(\#r,\omega)}

\def\calA{{\cal A}}
\def\calB{{\cal B}}

\def\PAmat{[\=P_\calA]}
\def\PBmat{[\=P_\calB]}

\begin{document}

\begin{center}

\LARGE{ {\bf  Electromagnetic surface waves at exceptional points
}}
\end{center}
\begin{center}
\vspace{10mm} \large

  \vspace{3mm}
  
 \textbf{Akhlesh  Lakhtakia}\footnote{E--mail: akhlesh@psu.edu}\\
 {\em NanoMM~---~Nanoengineered Metamaterials Group\\ Department of Engineering Science and Mechanics\\
Pennsylvania State University, University Park, PA 16802--6812, USA}\vspace{3mm}\\

 \textbf{Tom G. Mackay}\footnote{E--mail: tmackay@ed.ac.uk}\\
{\em School of Mathematics and
   Maxwell Institute for Mathematical Sciences\\
University of Edinburgh, Edinburgh EH9 3FD, UK}\\
and\\
 {\em NanoMM~---~Nanoengineered Metamaterials Group\\ Department of Engineering Science and Mechanics\\
Pennsylvania State University, University Park, PA 16802--6812,
USA}\vspace{3mm}\\

and\\
  
 \textbf{Chenzhang Zhou}\\
 {\em NanoMM~---~Nanoengineered Metamaterials Group\\ Department of Engineering Science and Mechanics\\
Pennsylvania State University, University Park, PA 16802--6812, USA}

\normalsize

\end{center}

\begin{center}
\vspace{15mm} {\bf Abstract}
\end{center}
Guided by the planar interface of two dissimilar linear, homogeneous mediums,
a Voigt surface wave arises due to an exceptional point of either of the two matrixes necessary to describe the 
spatial characteristics 
in the direction
normal to the planar interface. There is no requirement for either or both partnering mediums to be dissipative,
unlike   a Voigt plane wave
which can propagate only in a dissipative medium.

\section{Introduction}

Exceptional points \cite{Heiss} have gained considerable currency this century in the literature on
electromagnetics wave propagation in a passive linear homogeneous  medium. The planewave solution 
of the time-harmonic Maxwell  postulates  in this medium is given by
\begin{equation}
\label{eq1}
\Er = \#U_1 \exp\left(ik_1\ua\.\#r\right) + \#U_2 \exp\left(ik_2\ua\.\#r\right)\,,
\end{equation}
where $\#E\in\mathbb{C}^3$ is the electric field phasor, $\#r\in\mathbb{R}^3$ is the position vector, $\#U_1\in\mathbb{C}^3$ and
$\#U_2 \in\mathbb{C}^3$ are amplitude vectors, $i=\sqrt{-1}$,
$k_1\in\mathbb{C}$ and $k_2\in\mathbb{C}$ are  wavenumbers with non-negative
 imaginary parts,  the unit vector $\ua\in\mathbb{R}^3$ 
delineates the direction of propagation, and an $\exp(-i\omega t)$ dependence on time $t$ is implicit
with $\omega$ as the angular frequency.

However, Eq. \r{eq1} is not always applicable. Circumstances may arise
when $k_2=k_1$ and $\#U_1\parallel \#U_2$ \cite{Panch1958,Gerardin}, as was experimentally shown
by Voigt \cite{Voigt} in 1902 with biaxial absorbing dielectric mediums. There \textit{may} be
four isolated values of $\ua$ for which  Eq. \r{eq1} has to be replaced by \cite{Boyce,Borzdov}
\begin{equation}
\label{eq2}
\#E(\#r) = \les\#V_1   + i (\ua\.\#r)\#V_2\ris
 \exp\left(ik_1\ua\.\#r\right) \,,
\end{equation}
where $\#V_1\in\mathbb{C}^3$ and
$\#V_2 \in\mathbb{C}^3$. The   linear dependence on the distance
$\ua\.\#r$ along the direction of propagation in  Eq.~\r{eq2},
in addition to the usual exponential dependence common to both
Eqs.~\r{eq1} and \r{eq2}, is the hallmark of the \textit{Voigt plane wave} \cite{Panch1955-3,Grundmann,Brenier}.
Voigt plane waves cannot propagate in isotropic and uniaxial dielectric mediums, whether dissipative
or not; neither can  Voigt plane waves propagate in nondissipative biaxial dielectric mediums.

The specific directions of  propagation of a Voigt plane wave in a particular
medium are called the singular axes \cite{Voigt} of that medium. These axes
appear in band diagrams  as exceptional points \cite{Grundmann,Richter}. 
Exceptional points are readily appreciated through matrix algebra. Let the unit vector $\ub\in\mathbb{R}^3$
be orthogonal to $\ua$. Then, the unit vector $\uc=\ua\times\ub$ is orthogonal to both $\ua$ and $\ub$
and the triad $\lec\ua,\ub,\uc\ric$ forms a right-handed Cartesian coordinate system. The Maxwell divergence postulates
then yield $\ua\.\Br\equiv 0$ and $\ua\.\DDr\equiv 0$, where $\Br\in\mathbb{C}^3$ is the magnetic flux
density phasor and $\DDr\in\mathbb{C}^3$ is the electric displacement phasor. Planewave propagation can then be investigated
by forming the matrix ordinary differential equation
\begin{equation}
\label{eq3}
\frac{d}{d(\ua\.\#r)}
\les\begin{matrix}
\ub\.\#D \\ \uc\.\#D \\\ub\.\#B \\ \uc\.\#B
\end{matrix}\ris =
\les\=W\ris\.
\les\begin{matrix}
\ub\.\#D \\ \uc\.\#D\\ \ub\.\#B \\ \uc\.\#B
\end{matrix}\ris\,,
\end{equation}
where the 4$\times$4 matrix $\les\=W\ris$ is independent of $\#r$. 
Ordinarily but not always, $\les\=W\ris$ will have
four eigenvalues, each of algebraic multiplicity $1$ and geometric multiplicity $1$. The occurrence of
an exceptional point is  unambiguously
indicated by an eigenvalue of algebraic multiplicity $2$ and geometric multiplicity $1$.

Whereas previous research on exceptional points in electromagnetics has been largely confined
to planewave propagation in dissipative mediums, exceptional points are also relevant to surface-wave propagation, as
we have illustrated in the following sections of this Letter.  The propagation of an electromagnetic surface wave is
guided by the  planar interface of two  mediums $\calA$ and $\calB$ whose constitutive parameters
are invariant in any direction wholly contained in the interface plane \cite{Boardman,ESW_book}. 
Exceptional points can show
up when at least one of the two mediums is homogeneous. Salient features of the underlying theory 
of \textit{Voigt surface waves} are presented
in Sec.~\ref{theory}, with both mediums passive, linear, homogeneous, and bianisotropic \cite{EAB}.
Three examples are provided in Sec.~\ref{examples}.

The following notation is adopted:
 The permittivity and permeability of free space are denoted by $\epso$ and $\muo$, respectively. 
 The free-space wavenumber is written as  
 $\ko = \omega \sqrt{\epso \muo}$.   
Vectors are in boldface,
     $\lec \ux, \uy, \uz \ric$ is
the triad of unit vectors aligned with the Cartesian axes, and the position
vector $\#r =x_1\ux+x_2\uy+x_3\uz$. Dyadics are double underlined \c{Chen}.
Square brackets enclose matrixes and column vectors. The superscript ${}^T$ denotes the transpose.  The real and imaginary parts of complex quantities are delivered by the operators $\mbox{Re} \lec \cdot \ric$ and  $\mbox{Im} \lec \cdot \ric$, respectively.

\section{Theory} \label{theory}

In the canonical boundary-value problem for surface-wave propagation \cite{ESW_book}, 
medium $\calA$   occupies the half-space $x_3>0$ and medium $\calB$   the half-space $x_3<0$, their
interface being the plane $x_3=0$. 

Medium $\calA$ is taken to have the constitutive relations
\begin{equation}
\label{conA}
\left.\begin{array}{l}
\DDr =\epso\=\eps_\calA(\omega)\.\Er+\=\xi_\calA(\omega)\.\Hr
\\[5pt]
\Br =\=\zeta_\calA(\omega)\.\Er+\muo\=\mu_\calA(\omega)\.\Hr
\end{array}\right\}\,, \quad x_3>0\,,
\end{equation}
where $\=\eps_\calA$ is the relative permittivity dyadic, $\=\mu_\calA$ is
the relative permeability dyadic, and $\=\xi_\calA$ as well as $\=\zeta_\calA$
are the magnetoelectric dyadics. Medium $\calB$ has the analogous constitutive relations
\begin{equation}
\label{conB}
\left.\begin{array}{l}
\DDr =\epso\=\eps_\calB(\omega)\.\Er+\=\xi_\calB(\omega)\.\Hr
\\[5pt]
\Br =\=\zeta_\calB(\omega)\.\Er+\muo\=\mu_\calB(\omega)\.\Hr
\end{array}\right\}\,, \quad x_3<0\,.
\end{equation}
Henceforth, the dependences of various quantities
on $\omega$ are not explicitly identified.

The electromagnetic field phasors for surface-wave propagation 
are expressed everywhere as
\cite{ESW_book} 
\begin{equation} \label{planewave}
\left.\begin{array}{l}
 \#E (\#r)=  \les e_1(x_3)\ux + e_2(x_3)\uy+e_3(x_3)\uz \ris \, 
 \exp\les i q \le x_1 \cos \psi + x_2 \sin \psi \ri \ris \\[4pt]
 \#H  (\#r)=  \les h_1(x_3)\ux + h_2(x_3)\uy+h_3(x_3)\uz \ris \,
 \exp\les i q \le x_1 \cos \psi + x_2 \sin \psi \ri \ris 
 \end{array}\right\}\,,  \qquad x_3 \in(-\infty,\infty)\,,
\end{equation}
where ${q}\in\mathbb{C}$ is the surface  wavenumber. The angle $\psi\in\left[0,2\pi\right)$
specifies
 the direction of propagation in the $x_1$-$x_2$ plane, relative to  the $x_1$ axis.
 The phasor representations~\r{planewave}, when combined with
the source-free Faraday and Amp\'ere--Maxwell equations,  deliver 
 the 4$\times$4 matrix ordinary differential
equations \c{Billard,Berreman}
\begin{equation}
\label{MODE_A}
\frac{d}{dx_3}\les\#f  (x_3)\ris= \left\{
\begin{array}{ll}
i \PAmat\.\les\#f  (x_3)\ris\,,  \qquad   &x_3>0 \,,\vspace{8pt} \\
i \PBmat\.\les\#f  (x_3)\ris\,,  \qquad   &x_3<0\,,
\end{array}  \right.
\end{equation}
wherein
the  column 4-vector
\begin{equation}
\les\#f (x_3)\ris= 
\les
\begin{array}{c}%
e_ 1(x_3), \quad
e_2(x_3),\quad
h_1(x_3),\quad
h_2(x_3)
\end{array}
\ris^T
\label{f-def}
\end{equation} depends on $x_3$, but
the 4$\times$4  matrixes  $\PAmat$ and $\PBmat$ are independent of $x_3$.
Explicit expressions for $\PAmat$ and $\PBmat$ are available elsewhere \cite[Sec.~3.3.1]{ESW_book},
but are too cumbersome for repetition here.
The $x_1$-directed and $x_2$-directed components of the phasors
are algebraically connected to their   $x_3$-directed components  \c{EAB}.

\subsection{Half-space $x_3>0$}

\subsubsection{Ordinary surface waves}
Let us first consider the ordinary surface waves
for which  $\PAmat$ has four eigenvalues, 
each with algebraic multiplicity $1$ and geometric multiplicity $1$. 
Eigenvalues with negative  imaginary parts are irrelevant for  surface-wave propagation.
 Denoted
 by  $\alpha_{\calA 1}$ and $\alpha_{\calA 2}$,    the two eigenvalues appropriate for  surface-wave analysis are such that
 $\mbox{Im} \lec  \alpha_{\calA 1} \ric > 0$ and  $\mbox{Im} \lec  \alpha_{\calA 2} \ric > 0$.
 Explicit expressions for the corresponding eigenvectors $\les\#v_{\calA 1}\ris$
 and $\les\#v_{\calA 2}\ris$  can be derived by solving the equations
\begin{equation}
\le \PAmat - \alpha_{\calA  n} \les\=I\ris \ri \. \les\#v_{\calA n} \ris=\les \#0\ris\,,\qquad n\in\lec1,2\ric\,,
\end{equation}
where $\les\=I\ris$ is the 4$\times$4 identity matrix and $\les\#0\ris$ is the null column 4-vector.
The two remaining eigenvalues, $\alpha_{\calA 3}$ and $\alpha_{\calA 4}$, of $\PAmat$ are irrelevant   because
 $\mbox{Im} \lec  \alpha_{\calA 3} \ric < 0$ and  $\mbox{Im} \lec  \alpha_{\calA 4} \ric < 0$.

Thus the general solution of Eq.~\r{MODE_A}${}_1$  applicable
to ordinary surface waves that decay as $x_3 \to +\infty$ is given as 
 \begin{equation} \l{D_gen_sol}
\les\#f  (x_3)\ris =  C_{\calA 1}  \les\#v_{\calA 1}\ris  \exp \le i \alpha_{\calA 1 } x_3 \ri
+ C_{\calA 2}   \les\#v_{\calA 2} \ris  \exp \le i \alpha_{\calA 2} x_3 \ri
\,,\quad x_3 > 0\,.
\end{equation}
The complex-valued constants $C_{\calA 1}$ and $C_{\calA 2}$ herein are fixed 
by applying  boundary conditions at $x_3=0$. These boundary conditions involve
 \begin{equation} \l{bcD+}
\les\#f  (0^+)\ris =  C_{\calA 1}  \les\#v_{\calA 1}\ris   
+ C_{\calA 2}   \les\#v_{\calA 2} \ris  
\,.
\end{equation}

\subsubsection{Surface wave at an exceptional point}

Both eigenvalues with positive imaginary parts must be equal  at an exceptional point, i.e.,
$\alpha_{\calA 1} = \alpha_{\calA 2}=\alpha_\calA$, and the value of $q$ can be ascertained thereby.
The corresponding eigenvector $\les\#v_\calA\ris$ has to be determined first by solving
\begin{equation}
\le \les\=P_\calA\ris - \alpha_{\calA } \les\=I\ris \ri \. \les\#v_{\calA } \ris=\les \#0\ris\,,
\end{equation}
and a corresponding generalized eigenvector $\les\#w_\calA\ris$ has to be then determined by solving
 \cite{Boyce}
\begin{equation}
\le \les\=P_\calA\ris - \alpha_{\calA } \les\=I\ris \ri \. \les\#w_{\calA }\ris = \les\#v_{\calA }\ris\,.
\end{equation}
 
Thus, the general solution of Eq.~\r{MODE_A}${}_1$  representing a  surface wave that decays as $x_3 \to +\infty$
and holds at an exceptional point
can be stated as
\begin{equation} \l{DV_gen_sol}
\les\#f  (x_3)\ris = \Big( C_{\calA 1}  \les\#v_{\calA }\ris  + C_{\calA 2} \lec i  x_3 \, \les\#v_{\calA} \ris  + \les\#w_{\calA}\ris \ric  \Big) \exp \le i \alpha_{\calA } x_3 \ri\,,\quad x_3 > 0\,.
\end{equation}
The complex-valued constants $C_{\calA 1}$ and $C_{\calA 2}$ herein are fixed 
by applying  boundary conditions at $x_3=0$. These boundary conditions involve
 \begin{equation} \l{bcDV+}
\les\#f  (0^+)\ris =  C_{\calA 1}  \les\#v_{\calA }\ris   
+ C_{\calA 2}   \les\#w_{\calA } \ris  
\,.
\end{equation}

\subsection{Half-space $x_3<0$}
Equation~\r{MODE_A}${}_2$ has to be solved in the same way as Eq.~\r{MODE_A}${}_1$. The
matrix $\PBmat$ has $\alpha_{\calB n}$ and $[\#v_{\calB n}]$, $n\in\left[1,4\right]$, as its $n$th eigenvalue
and eigenvector, respectively. Without loss of generality, we assume  that each of the four eigenvalues of $\PBmat$
has algebraic multiplicity $1$ and geometric multiplicity $1$.
After labeling the eigenvalues   such that
$\mbox{Im}\lec {\alpha_{\calB 3}} \ric<0$ and $\mbox{Im}\lec {\alpha_{\calB 4}} \ric<0$,
we set  
\begin{equation}
[\#f(0^-)]= C_{\calB 3}  \les\#v_{\calB 3}\ris   
+ C_{\calB 4}   \les\#v_{\calB 4} \ris 
\end{equation}
for surface-wave propagation,
where the complex-valued constants $C_{\calB 3}$ and $C_{\calB 4}$ are fixed 
by applying  boundary conditions at $x_3=0$.
The other two eigenvalues of  $[\=P_\calB]$ pertain to waves that amplify as $x_3\to-\infty$
and cannot therefore contribute to the surface wave.

\subsection{Dispersion equation}

 The continuity of the tangential  components of the electric and magnetic field
 phasors across the interface plane $x_3=0$ imposes
 four boundary conditions that are represented compactly as
 \begin{equation}
 \label{2.23-AL}
 \les\#f(0^+)\ris=  \les\#f(0^-)\ris
 \,.
 \end{equation}
Accordingly,
\begin{equation}
\les \=Y \ris \. \les \:
 C_{\calA 1}, \quad
  C_{\calA 2}, \quad
   C_{\calB 3}, \quad
    C_{\calB 4} \:
 \ris^T =  \les\#0\ris\,,
\end{equation}
wherein the 4$\times$4 characteristic matrix $\les \=Y \ris$ must be singular for  surface-wave propagation \c{ESW_book}.
The dispersion equation 
\begin{equation}
\l{dispersion_eq}
\left\vert \les \=Y \ris\right\vert = 0,
 \end{equation}  can be numerically solved for $q$ for a fixed value
 of $\psi$, by the Newton--Raphson method \c{N-R} for example.

\section{Illustrative Examples}\label{examples}
Examples of electromagnetic 
surface waves corresponding to the exceptional points of  $\PAmat$ have recently become available 
\c{MZL_PRSA,ZML_JOSAB,ZML_PRA},
though the exceptional nature of those surface waves has not been demonstrated yet. In order to
highlight the exceptional nature of the occurrence of a
Voigt surface wave, we now present three examples. For all these
examples, we have set $\=\xi_{\calA}=\=\xi_{\calB}=\=0$, $\=\zeta_{\calA}=\=\zeta_{\calB}=\=0$,
and $\=\mu_{\calA}=\=\mu_{\calB}=\=I$, where $\=0$ is the null dyadic and $\=I=\ux\ux+\uy\uy+\uz\uz$
is the identity dyadic.

 \begin{figure}[!htb]
\centering
\begin{subfigure}{0.49\textwidth}
 \includegraphics[width=7.3cm]{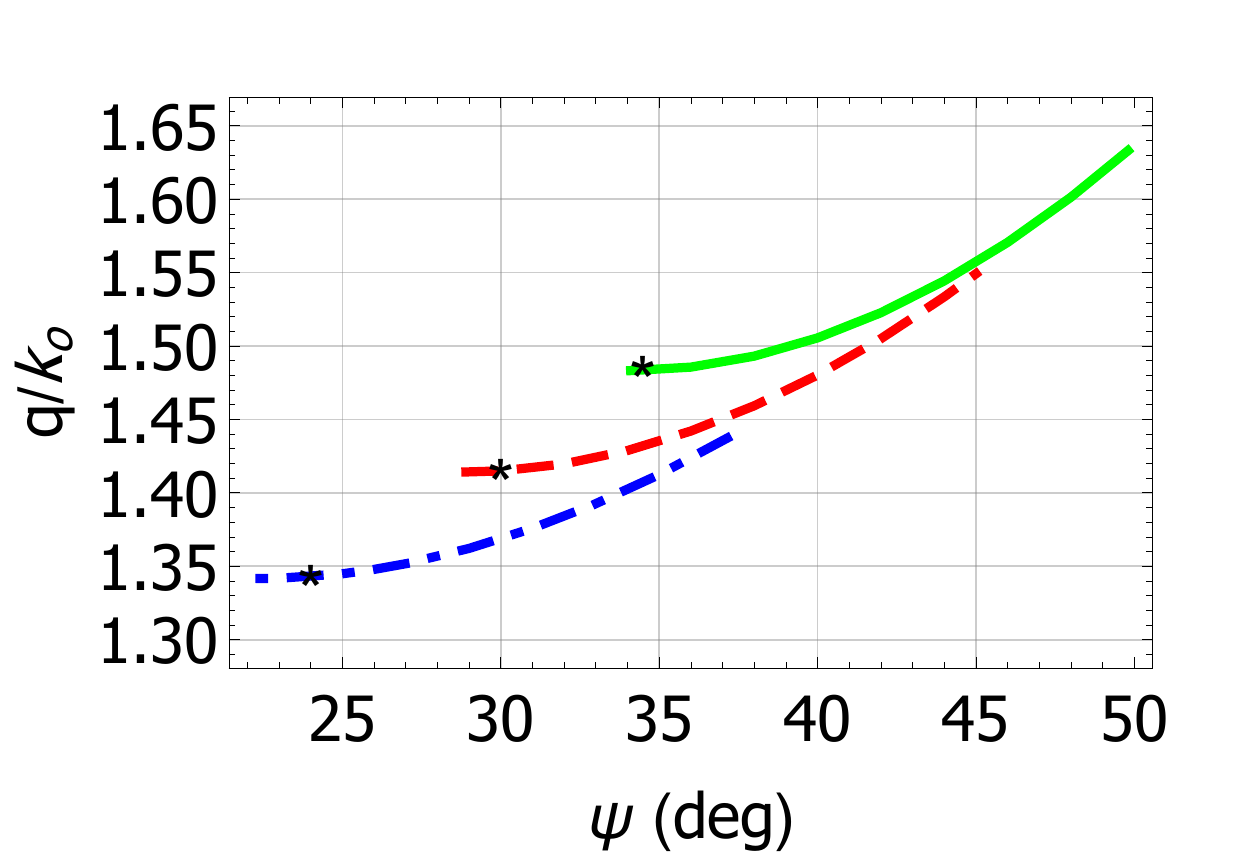} 
 \caption{}
 \end{subfigure} \hfill
 \begin{subfigure}{0.49\textwidth}
  \includegraphics[width=7.3cm]{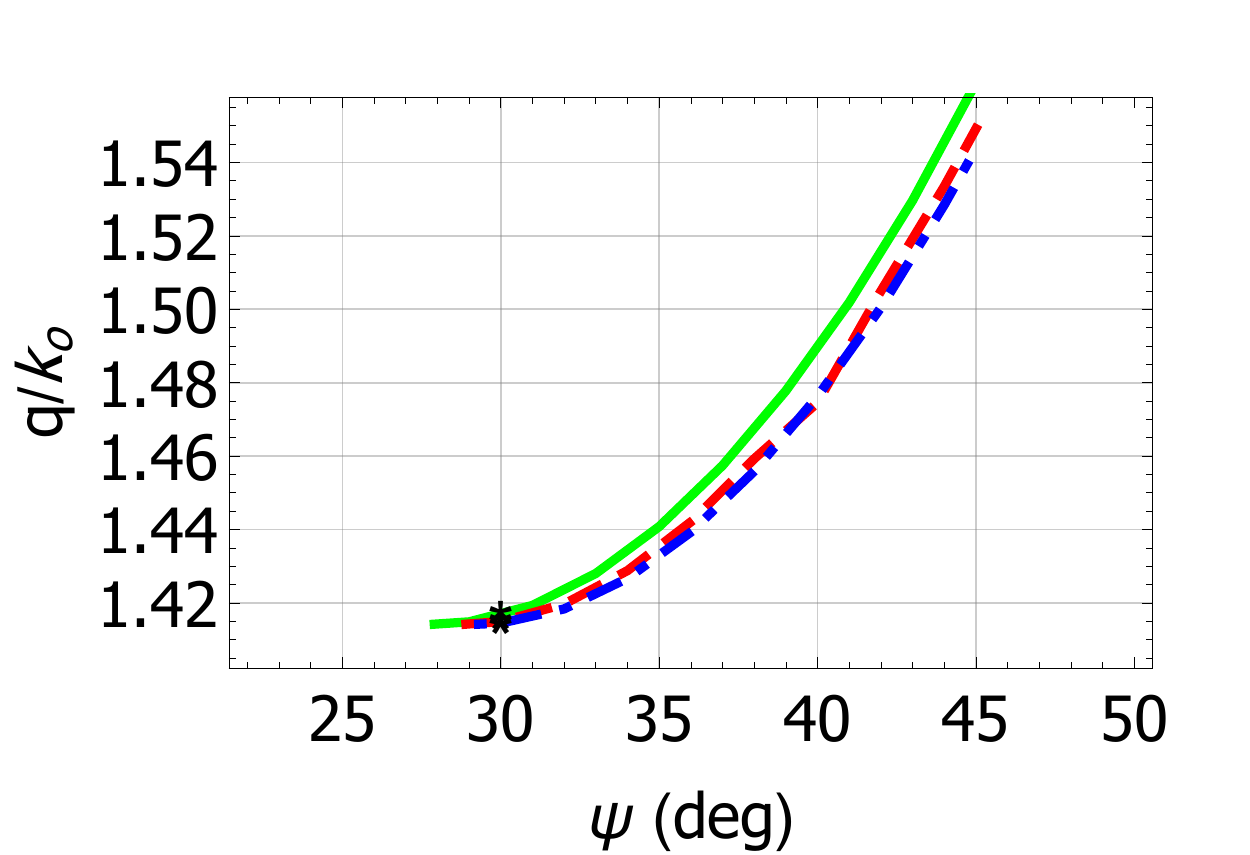}
 \caption{}
  \end{subfigure}\\
  \begin{subfigure}{0.49\textwidth}
     \includegraphics[width=7.3cm]{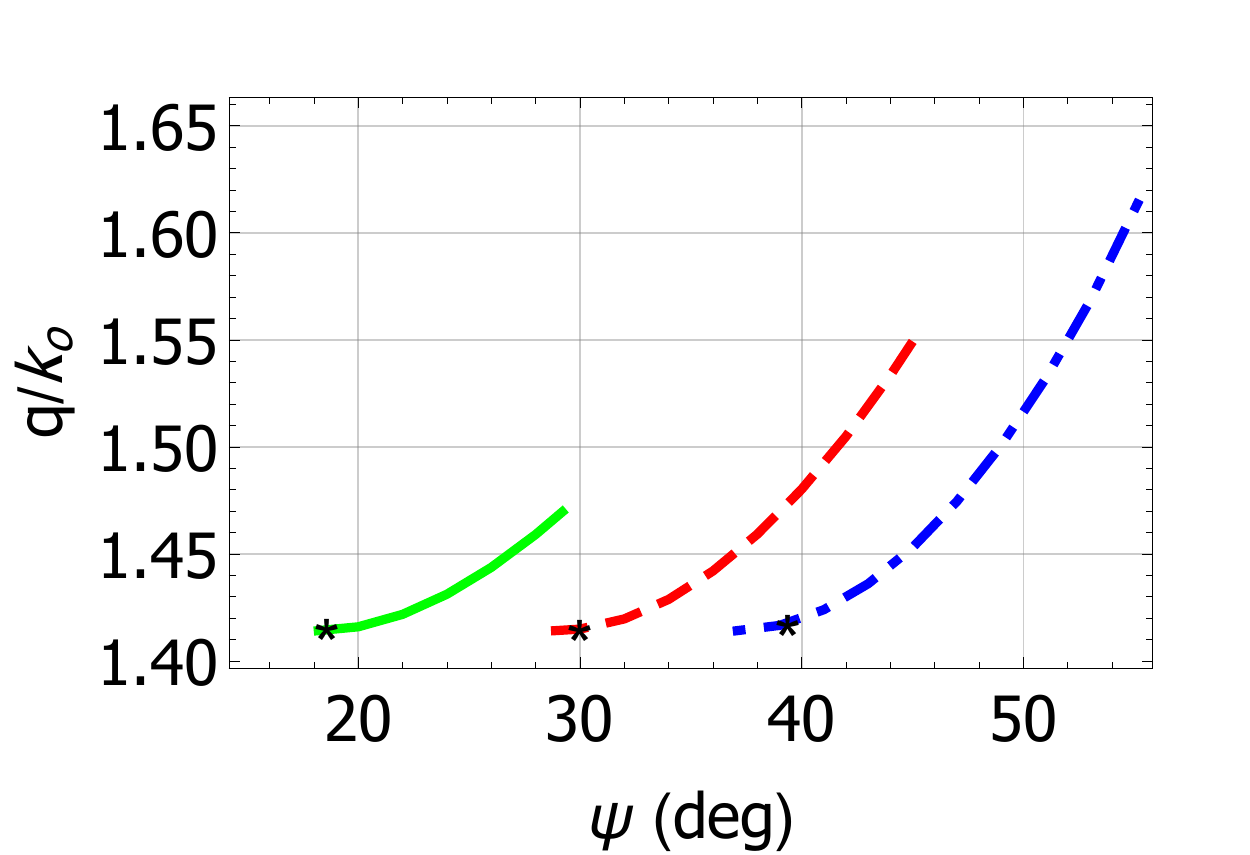} 
     \caption{}
     \end{subfigure}
 \caption{\label{DyakonovFigure1} 
Plots of  $q/\ko $
versus 
$\psi $ 
for Dyakonov surface waves guided by the planar interface of a non-dissipative uniaxial dielectric medium and a non-dissipative isotropic dielectric
medium described by Eqs.~\r{uni-iso}.
 (a)  $\eps^s_\calA=1.5, \eps^t_\calA=6$, and $\eps_\calB = 1.8$ (blue broken-dashed curve), $ 2$ (red dashed curve), $2.2$ (green solid curve);
 (b)  $\eps^s_\calA=1.5, \eps_\calB=2$, and $\eps^t_\calB = 5.5$ (blue broken-dashed curve), $ 6$ (red dashed curve), $6.5$ (green solid curve);
 and
 (c)  $\eps^t_\calA=6, \eps_\calB=2$, and $\eps^s_\calA = 1.2$ (blue broken-dashed curve), $ 1.5$ (red dashed curve), $1.8$ (green  solid curve). 
A  Dyakonov--Voigt surface wave corresponding to an exceptional point of $\les{\underline{\underline P}}_\calA\ris$ is identified by a black star in each curve.
    }
\end{figure}

We chose the relative permittivity dyadic of medium $\calA$ to be  uniaxial with its preferred   axis parallel to $\ux$, 
and the relative permittivity dyadic of
medium $\calB$ to represent a scalar medium \c{MZL_PRSA} for all three examples; hence,
\begin{equation}
\left.\begin{array}{l}
\=\eps_\mathcal{A} = \eps_\mathcal{A}^{\rm s} \=I + \le
\eps_\mathcal{A}^{\rm t} - \eps_\mathcal{A}^{\rm s} \ri \,
\ux   \ux
\\[6pt]
\=\eps_\mathcal{B}  =\eps_\mathcal{B}\=I
\end{array}\right\}\,.
\label{uni-iso}
\end{equation} 
For these  constitutive relations,
if a  surface wave   exists for angle $\psi = \psi^\star$, then
surface-wave propagation is also possible for $\psi = - \psi^\star$ and $\psi = \pi \pm \psi^\star$.

\subsection{Example No. 1}\label{ex1}

For the first example, we chose $\eps_\mathcal{A}^{\rm s}\in\mathbb{R}$, $\eps_\mathcal{A}^{\rm t}\in\mathbb{R}$,
and $\eps_\mathcal{B}\in\mathbb{R}$. Furthermore, we chose all three constitutive parameters to be positive. 
Ordinary surface waves guided by the planar interface of the chosen mediums are classified
as Dyakonov surface waves \c{MSS,Dyakonov88,Takayama_exp}. 

Figure~\ref{DyakonovFigure1} presents plots of the normalized wavenumber $q/\ko$ versus $\psi$
found for the quadrant $0\leq\psi\leq\pi/2$ when all three constitutive parameters
$\eps_\mathcal{A}^{\rm s}$, $\eps_\mathcal{A}^{\rm t}$, and $\eps_\mathcal{B}$ are real and positive.
Despite both partnering mediums $\calA$ and $\calB$ being non-dissipative, a \textit{solitary} point 
 on each curve in Fig.~\ref{DyakonovFigure1} is the manifestation of
an exceptional point of $\PAmat$. The surface wave corresponding to this exceptional point is classified
as a Dyakonov--Voigt surface wave \cite{MZL_PRSA}.

 The values of
$\eps_\mathcal{B}$,
 $\eps_\mathcal{A}^{\rm t}$, and  $\eps_\mathcal{A}^{\rm s}$   were varied for
Figs.~\ref{DyakonovFigure1}(a), (b), and (c), respectively.
While the angle of propagation $\psi$ for the Dyakonov--Voigt surface wave is highly sensitive to the values of $\eps_\mathcal{A}^{\rm s}$
and $\eps_\mathcal{B}$ as we see in Figs.~\ref{DyakonovFigure1}(a) and (c), the same is not true of  $\eps_\mathcal{A}^{\rm t}$ as can be observed in
 Fig.~\ref{DyakonovFigure1}(b).
 Also, while the surface wavenumber $q$ for the Dyakonov--Voigt surface wave is highly sensitive to the values of 
 $\eps_\mathcal{B}$ as we see in Fig.~\ref{DyakonovFigure1}(a), the same is not true of 
 $\eps_\mathcal{A}^{\rm s}$ and
  $\eps_\mathcal{A}^{\rm t}$ as we note from
 Figs.~\ref{DyakonovFigure1}(b) and (c).

 \begin{figure}[!htb]
\centering
\begin{subfigure}{0.49\textwidth}
 \includegraphics[width=7.3cm]{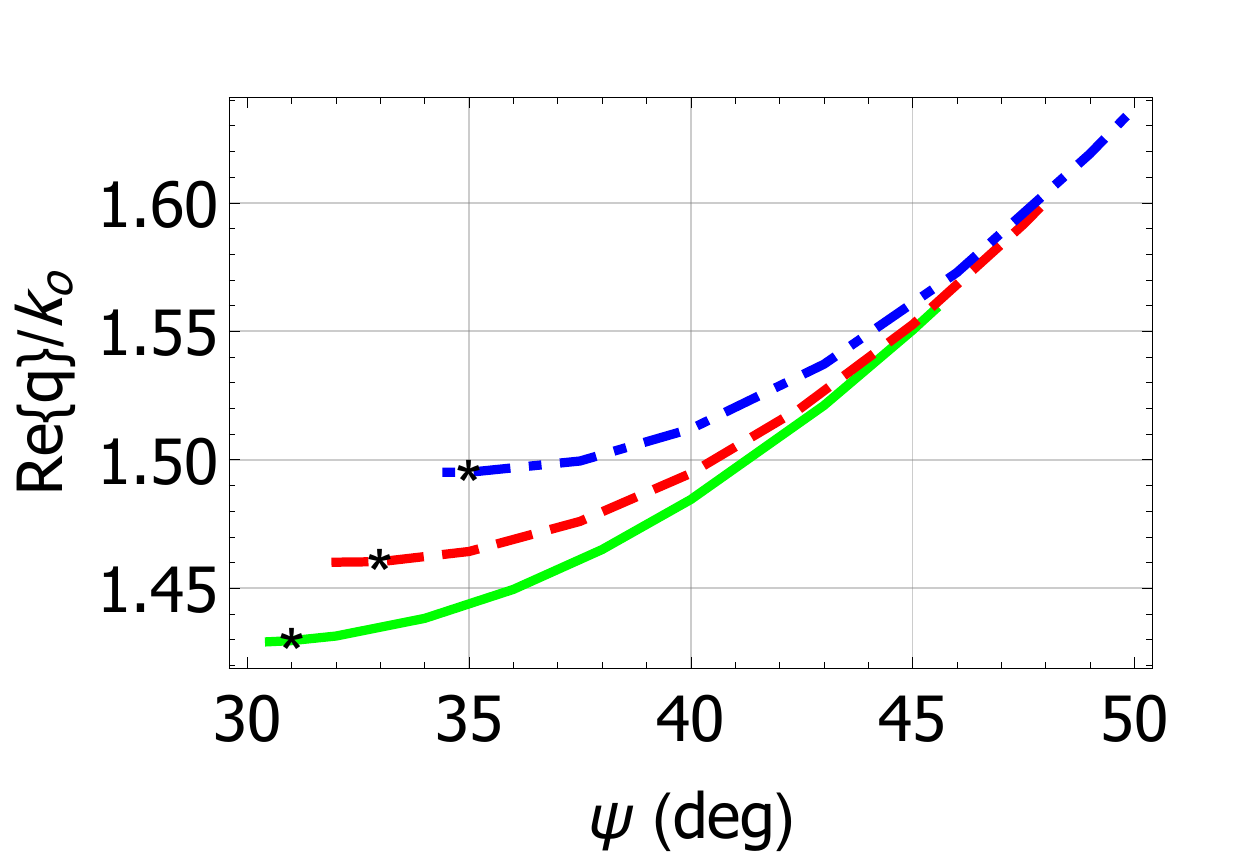} 
 \caption{}
 \end{subfigure} \hfill
 \begin{subfigure}{0.49\textwidth}
  \includegraphics[width=7.3cm]{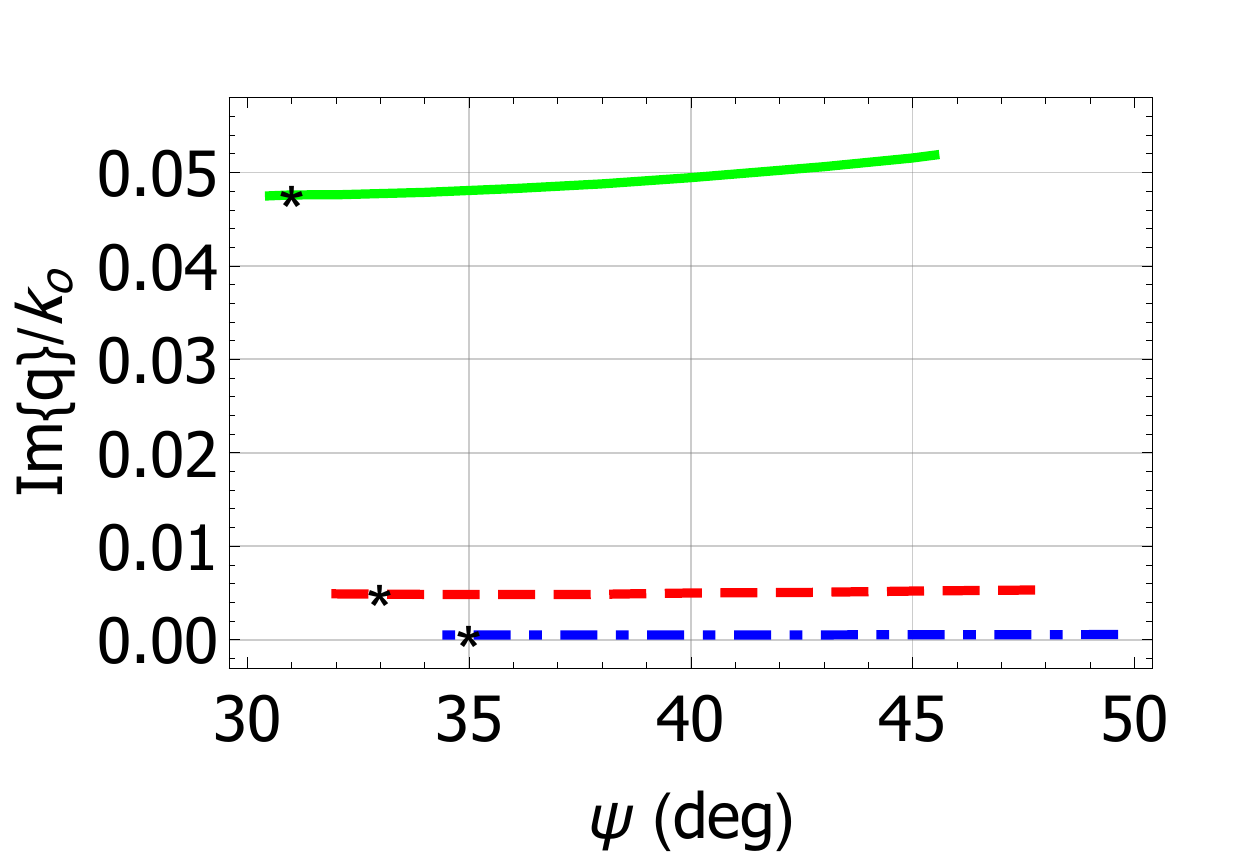}
 \caption{}
  \end{subfigure}\\
 \caption{\label{DyakonovFigure2} 
Plots of  (a) ${\rm Re}\lec{q}\ric/\ko $ and (b) ${\rm Im}\lec{q}\ric/\ko $ 
versus 
$\psi $ 
for Dyakonov surface waves guided by the planar interface of a dissipative uniaxial dielectric medium and a dissipative isotropic dielectric
medium described by Eqs.~\r{uni-iso}.
  $\eps^s_\calA=1.5+\delta i $ and  $\eps^t_\calA=6+4\delta i$ with  $\delta=0.1$, $\eps_\calB=2.0398 + 0.1360 i$ (green solid curves);  $\delta=0.01$, $\eps_\calB=2.1320 + 0.0142 i$ (red dashed curves); and  $\delta=0.001$, $\eps_\calB=2.2354 + 0.0015 i$ (blue broken-dashed curves). On each curve,
 a  {Dyakonov--Voigt} surface wave corresponding to an exceptional point of $\les{\underline{\underline P}}_\calA\ris$ is identified by a black star.
    }
\end{figure}

\subsection{Example No. 2}\label{ex2}

Whereas both partnering mediums were chosen to be non-dissipative for Fig.~\ref{DyakonovFigure1}, both
were chosen to be dissipative  for the second example, i.e., $\eps_\mathcal{A}^{\rm s}\in\mathbb{C}$, $\eps_\mathcal{A}^{\rm t} \in\mathbb{C}$, and $\eps_\mathcal{B}\in\mathbb{C}$. Also, both the real and the imaginary parts
of every one of these three constitutive parameters were chosen to be positive. 

Plots of ${\rm Re}\lec{q}\ric/\ko$ and ${\rm Im}\lec{q}\ric/\ko$ versus $\psi$
found for the quadrant $0\leq\psi\leq\pi/2$ are shown in Fig.~\ref{DyakonovFigure2}. 
 Each curve in this figure has a solitary
solution that is  identified by a black star. Every star represents a Dyakonov--Voigt surface wave engendered by an
exceptional point of $\PAmat$ \c{ZML_JOSAB}. 

The degrees of dissipation of the partnering materials, as well as $\mbox{Re}
\lec
\eps_\calB \ric$, were varied for Fig.~\ref{DyakonovFigure2}.
The angle of propagation $\psi$ and the real and imaginary parts of the  surface wavenumber $q$ for the Dyakonov--Voigt surface waves
represented in  Fig.~\ref{DyakonovFigure2} are
  highly sensitive to the degrees of dissipation of the partnering materials and  $\mbox{Re} \lec \eps_\calB \ric$.

 \begin{figure}[!htb]
\centering
\begin{subfigure}{0.49\textwidth}
 \includegraphics[width=7.3cm]{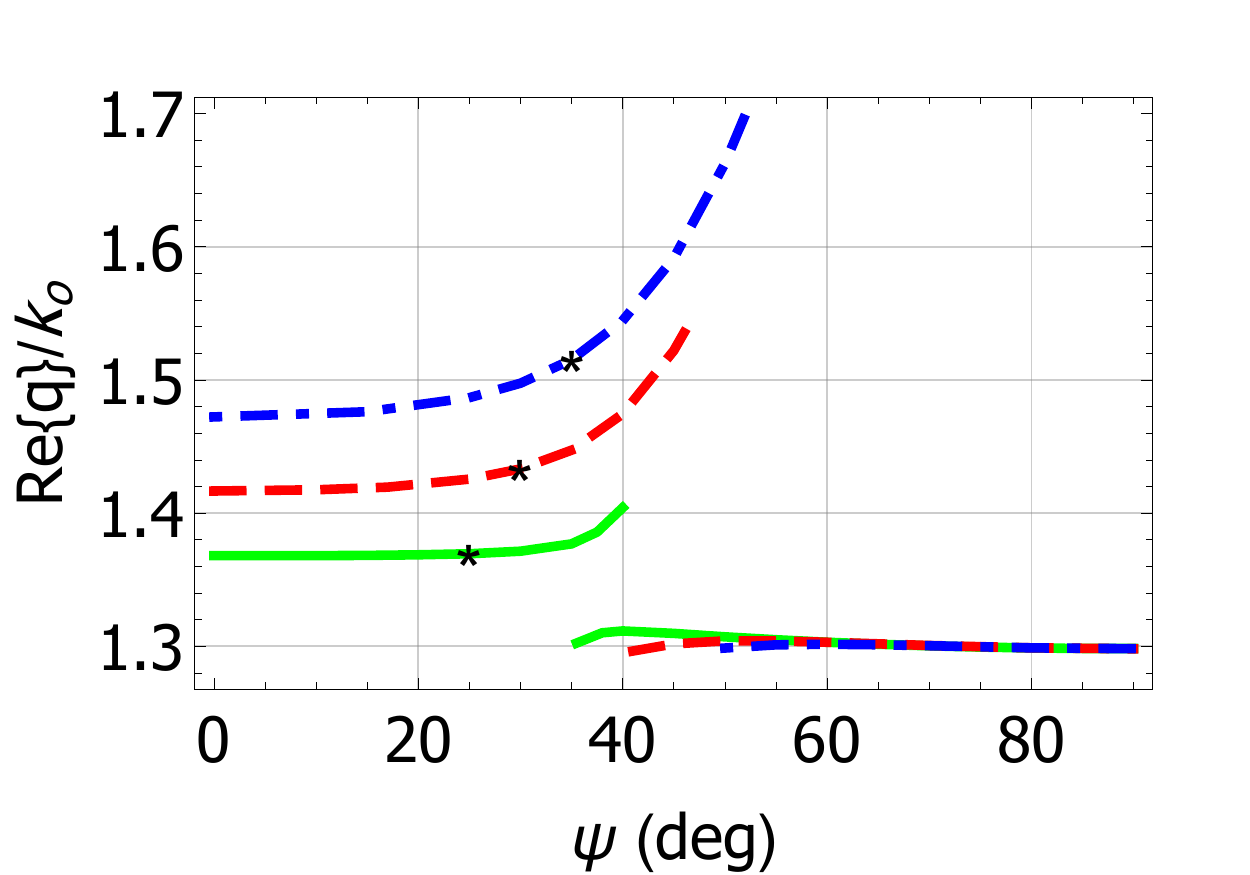} 
 \caption{}
 \end{subfigure} \hfill
 \begin{subfigure}{0.49\textwidth}
  \includegraphics[width=7.3cm]{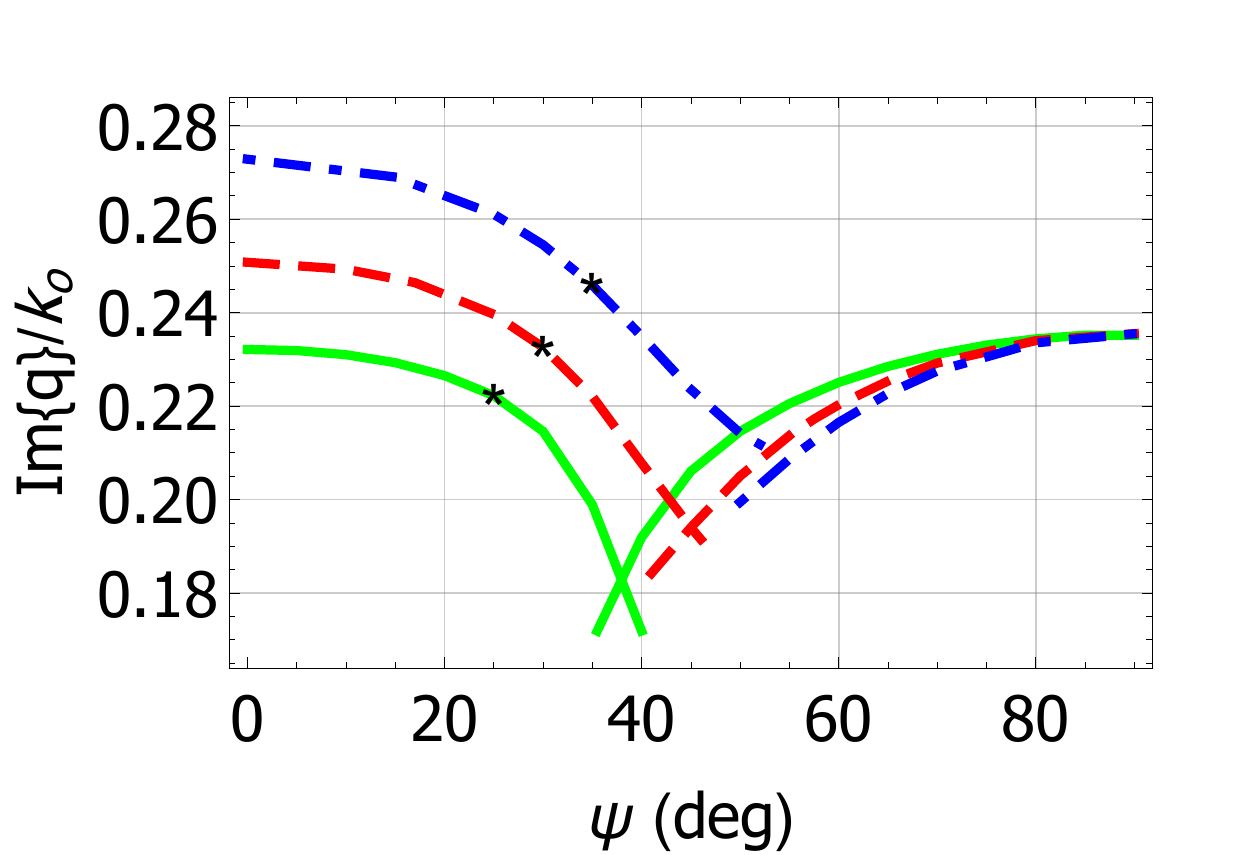}
 \caption{}
  \end{subfigure}\\

 \caption{\label{SPPFigure}  Plots of  ${\rm Re}\lec{q}\ric/\ko $ and ${\rm Im}\lec{q}\ric/\ko $
versus 
$\psi $ 
for surface-plasmon-polariton  waves guided by the planar interface of a dissipative uniaxial dielectric medium and an isotropic plasmonic
medium described by Eqs.~\r{uni-iso}. $\eps_\calB=-16.07+0.44 i$ and $\eps^s_\calA=1.5+0.5 i$ with $\eps^t_\calA=3.1559 + 0.0459 i$ (green solid curves), $4.4516 + 0.2285 i$ (red dashed curves), $5.9760 + 0.4572 i$ (blue broken-dashed curves).
A  surface-plasmon-polariton wave corresponding to an exceptional point of $\les{\underline{\underline P}}_\calA\ris$ is identified by a black star.
    }
\end{figure}

\subsection{Example No. 3}\label{ex3}
For the third and final example, we again chose
$\eps_\mathcal{A}^{\rm s}\in\mathbb{C}$, $\eps_\mathcal{A}^{\rm t} \in\mathbb{C}$, and $\eps_\mathcal{B}\in\mathbb{C}$. 
With ${\rm Re}\lec\eps^s_\calA\ric > 0$, ${\rm Im}\lec\eps^s_\calA\ric > 0$, 
${\rm Re}\lec\eps^t_\calA\ric > 0$, and ${\rm Im}\lec\eps^t_\calA\ric > 0$, medium $\calA$ is dissipative dielectric; however,
 with ${\rm Re}\lec\eps_\calB\ric < 0$ and ${\rm Im}\lec\eps_\calB\ric > 0$, medium $\calB$ is plasmonic. Ordinary
 surface waves guided by the planar interface of these two mediums are called surface-plasmon-polariton waves \cite{Sprokel,Depine97}.

Plots of ${\rm Re}\lec{q}\ric/\ko$ and ${\rm Im}\lec{q}\ric/\ko$ versus $\psi$
found for the quadrant $0\leq\psi\leq\pi/2$ are shown in Fig.~\ref{SPPFigure}.  
Surface-plasmon-polariton waves exist for all $\psi \in \les 0, \pi/2 \ris$, unlike the Dyakonov surface waves represented in Figs.~\ref{DyakonovFigure1} and \ref{DyakonovFigure2} which have smaller angular existence domains. For each set of values of $\lec \eps^s_\calA, \eps^t_\calA, \eps_\calB \ric$, a pair of 
surface-plasmon-polariton
solution curves exist. For each pair of curves, there is an interval of midrange $\psi$ values for which the curves overlap. Thus, for propagation directions specified by midrange values of $\psi$, two  surface-plasmon-polariton waves exist for fixed values of $\lec \eps^s_\calA, \eps^t_\calA, \eps_\calB \ric$ \c{ZML_PRA}.
For each pair of solution curves corresponding to a fixed set of values of $\lec \eps^s_\calA, \eps^t_\calA, \eps_\calB \ric$, there is 
a solitary
solution,  identified by a black star. Every star represents a surface-plasmon-polariton-Voigt   wave engendered by an
exceptional point of $\PAmat$ \c{ZML_PRA}.

\section{Concluding Remarks}
Exceptional points have been generally associated in electromagnetics  with plane waves propagating in
a linear, homogeneous, dissipative, biaxial dielectric medium that fills up a
certain region. Here, we have extended the scope of exceptional points to affect
surface-wave propagation guided by the planar interface of two different linear, homogeneous, bianisotropic mediums.

Section~\ref{examples} demonstrates that
there is no reason for either or both partnering mediums to be dissipative, for a surface wave to be engendered
by an exceptional point of either of the two matrixes necessary to describe the spatial characteristics in the direction
normal to the planar interface. This attribute of a Voigt surface wave is in marked contrast to that of a Voigt plane wave,
which can propagate only in a dissipative medium (or an active medium \c{Voigt_active}). What seems essential for a Voigt wave, whether a plane wave
or a surface wave, is that it must decay in some direction. And, in the direction of decay, it must have a spatial variation
that is the product of a linear function and an exponential function.

 \vspace{5mm}

\noindent {\bf Acknowledgments.}
This work was supported  in part by
 US NSF (grant number DMS-1619901) and EPSRC (grant number EP/S00033X/1).
AL thanks the Charles Godfrey Binder Endowment at the Pennsylvania State University    for partial support of his research endeavors.

\end{document}